\documentstyle[12pt]{article}

\def\lapp{{\ \lower 0.6ex \hbox{$\buildrel<\over\sim$}\ }}
\def\gapp{{\ \lower 0.6ex \hbox{$\buildrel>\over\sim$}\ }}
\def\lumi{\int{\cal L}}
\def\invpb{{\rm pb}^{-1}}
\def\GeV{{\rm GeV}}
\def\decomp{1}
\def\sigmaww{2}
\def\ratios{3}
\def\fivemasses{4}
\def\statsens{5}
\def\multsens{6}
\def\syssens{7}
\def\widthvar{8}
\def\lepfourj{9}
\def\lepiifourj{10}
\begin{document}
\begin{titlepage}
\vspace*{-1cm}
\begin{flushright}
DTP/95/24 \\
February 1995
\end{flushright}
\vskip 1.cm
\begin{center}
{\Large\bf
The Measurement of M$_W$ from the W$^+$W$^-$ \\[2mm]
Threshold Cross Section at LEP2}
\vskip 1.cm
{\large W.J.~Stirling}
\vskip .4cm
{\it Departments of Physics and Mathematical Sciences, University of Durham \\
Durham DH1 3LE, England }\\
\vskip 1cm
\end{center}
\begin{abstract}
The rapid increase of the  $e^+e^- \to W^+W^-$ cross section
in the threshold region provides a method for measuring $M_W$
at LEP2.  The dependence of the theoretical cross section --- including the
effects
of the finite $W$ width, Coulomb interactions and initial state
radiation --- on the collider energy and on the mass and width of the
$W$ is investigated in order to deduce the optimal collider energy
which maximizes the sensitivity to  $M_W$.
Statistical and systematic uncertainties are studied in detail.
The background contribution from QCD four-jet production is
estimated.
It is shown that by running  LEP2 at $\sqrt{s} \approx 161$~GeV
and obtaining a luminosity of order 100~pb$^{-1}$ per experiment,
an uncertainty on the measured value of $M_W$
of   $\pm 70$~MeV should be achievable.
\end{abstract}
\vfill
\end{titlepage}
\newpage

\section{Introduction}

A precision measurement of the mass of the $W$ boson
is one of the most important physics goals of the LEP2
$e^+e^-$ collider. The impact of such a measurement
on tests  of the  Standard Model
 and constraints on new physics beyond is well documented,
see for example Sec.~10 of Ref.~\cite{AACHEN} or Sec.~8
of Ref.~\cite{DPF}.

The current `world average' value is \cite{CDFWMASS}
\begin{equation}
M_W = 80.23 \pm 0.18\ \GeV          \; ,
\label{currentmw}
\end{equation}
obtained by combining the results from the three $p \bar p$ collider
experiments: CDF, D0 and UA2. The precision will improve with increasing
luminosity at the Tevatron collider. In fact it is likely that the error
will decrease to $O(100)$~MeV in the next few years, and may
ultimately be as small as $50$~MeV \cite{DPF}. Clearly the aim of LEP2
should be to at least match this level of precision.

At least three methods of measuring $M_W$ at LEP2 have been proposed,
see for example Refs.~\cite{AACHEN,DELPHI}. The most direct method,
and at first sight the most promising, is to reconstruct $M_W$
directly from its hadronic decay products using the decay channels
\begin{eqnarray}
\label{eq:qqqq}
W^+W^- &\to & q \bar{q} q \bar{q} \; , \\
W^+W^- &\to & q \bar{q} \ell  \nu \; .
\end{eqnarray}
If, for example, an integrated luminosity of $\lumi \simeq 500$~pb$^{-1}$
 can be obtained
at a collider energy in the range 170$-$200~GeV, then the statistical
uncertainty for this method (combining four experiments) is well
below 50~MeV \cite{DELPHI}. Notwithstanding the difficulties of reconstructing
and measuring hadronic jets, the systematic errors appear to be
under control as well.  There is, however, one important caveat.
It has recently been argued \cite{RECONNECT} that
 `colour reconnection' effects ---
strong interactions between the separate hadronic final states
of each $W$ in (\ref{eq:qqqq}) --- {\it may} distort the mass measurement.
The studies which yield small systematic errors on $M_W$ from reconstructing
the four-jet final state  assume that
the $W^+W^-$ system can be treated as a superposition of two
separate $W$ decays, an approximation which may not be valid
at the necessary level of precision.

A second method of measuring $M_W$ is to use the fact that the end-points
of the lepton spectrum in $W^+W^- \to  q \bar{q} \ell  \nu$ depends
quite sensitively on the  $W$ mass:
\begin{equation}
E_-  \leq E_{\ell}  \leq E_+\; , \qquad
 E_\pm =  {\sqrt{s}\over 4}\; \left( 1 \pm \sqrt{1-4M_W^2/s} \right) \; .
\end{equation}
The problem with this method is that the statistical error is large
for the total luminosity envisaged at LEP2: the net branching ratio
for the $q \bar q \ell \nu$  final
state ($ \approx 8/27 $, for $\ell=e,\mu$) is quite small, and only
a fraction of
events close to the end-points are sensitive to $M_W$. In Ref.~\cite{DELPHI}
a statistical error on $M_W$ of order 300~MeV was obtained using
this method.

The third method, the subject of the present work,  is to use the rapid
increase
of the $W^+W^-$ cross section at $\sqrt{s} \sim 2 M_W$ to measure the mass.
Although the rise of the Born (on-shell $W$) cross section is quite
sharp, $\sigma \sim \beta$ where $\beta^2 = 1 - 4 M_W^2/s$,
various effects, for example the finite $W$ width,
smear out the threshold cross section, and so in practice the procedure
is to fit a theoretical expression for the cross section, with $M_W$
a free parameter, to the data.
The advantage of this method is that it is a direct `kinematic'
measurement of the $W$ mass, in the sense that the  dependence of
the cross section on $M_W$ near threshold comes mainly from the
opening up of the phase space
rather than from the matrix element.
There are, however, several important questions to address: (i) is it necessary
to run at several collider energes to scan the threshold region, or is
it better to run at a single `optimal' energy, (ii) what is the resulting
statistical precision on the mass measurement, (iii) how accurately
must the cross section be measured, and how precisely does the theoretical
cross section have to be known,  to keep the systematic uncertainties
well below the statistical uncertainty?

In this paper we attempt to answer these questions. We first of all
discuss in some detail the status of the theoretical calculation
of the $e^+e^-\to W^+W^-$ cross section in the threshold region. We
derive and use an expression which, we believe, contains the bulk
of the higher-order effects. We then study the variation of this cross section
with $\sqrt{s}$, $M_W$, $\Gamma_W$, ... to estimate
 the optimal collider energies, and the corresponding
statistical and systematic uncertainties.

There have been several other recent studies along broadly similar lines.
In the theoretical studies of
Refs.~\cite{CAHN,CHEN}, the emphasis was on estimating the
uncertainty on $M_W$ as a function of the collider energy.
 More recently, ongoing
studies within the LEP collaborations have also been reported
\cite{GRUEN,KATSAN,VALASSI}. These address additional issues such as
the problem of background contributions to the cross section.
Where these studies overlap with the present work, there is a large
measure of agreement. Earlier  work
\cite{AOKI,PECCEI,HAGIWARA,HIOKI,KUHN} focused mainly
on theoretical issues concerning the calculation of the  $W^+W^-$
cross section at LEP2 energies, and also of the impact on the
cross section of new physics beyond the Standard Model.

The present work is closest in spirit to Ref.~\cite{CAHN}, in which many of the
concepts to be discussed below were first presented. However our
study extends the analysis of Ref.~\cite{CAHN} in several important ways.
We include, for example, the leading Coulomb corrections which
have a significant effect on the threshold cross section, and consequently
on the extracted value of $M_W$. We also attempt to quantify the
most important background cross section, and assess its contribution
to the systematic uncertainty on the $W$ mass.

\section{Calculation of the $W^+W^-$ cross section}

It is not our intention
to perform  a complete  calculation of the theoretical
 cross section including,  for example, all next-to-leading order
 electroweak corrections or all background contributions. In fact
 the  next-to-leading order electroweak corrections for four fermion
 (i.e. off-shell $W^+W^-$) production are only partially known at present,
see for example Ref.~\cite{BEENDENNER} for a detailed and up-to-date review.
Our aim instead is to use a theoretical cross section which includes
 the most important higher-order effects to quantify the dependence
 on $M_W$. We will then be able to estimate the effect on the
 extracted $M_W$ value of subsequent small variations in the theoretical
 prediction.

We begin by writing the cross section, schematically,\footnote{We note
that this decomposition of the cross section into
 `signal' and `background' contributions is practical rather than
theoretically rigorous, since neither contribution is separately gauge
invariant nor experimentally distinguishable in general.} as
\begin{eqnarray}
\sigma &= & \sigma_{\rm WW} \; +\; \sigma_{\rm bkd} \; , \nonumber \\
\sigma_{\rm WW} &= & \sigma_0^{\rm WW} \; (1 + \delta_{\rm EW}
+ \delta_{\rm QCD}) \; ,
\end{eqnarray}
where the various terms correspond to
\begin{itemize}
\item[{(i)}] $\sigma_0^{\rm WW}$: the Born contribution from the 3
leading-order
diagrams for $e^+e^- \to W^+W^-$ involving $t$-channel $\nu$ exchange
and $s$-channel $\gamma$ and $Z$ exchange, calculated using off-shell $W$
propagators.
\item[{(ii)}] $\delta_{\rm EW}$: higher-order electroweak radiative
corrections,
including loop corrections, real photon emission, etc.
\item[{(iii)}] $\delta_{\rm QCD}$: higher-order QCD corrections
to $W^+W^-$  final states containing $q \bar q$ pairs. Since we are
only concerned with the {\it total} cross section in this study,
the  effect of these is to generate small
corrections  to the hadronic branching ratios which are entirely
straightforward to calculate and take into account.
\item[{(iv)}] $\sigma_{\rm bkd}$: `background' contributions, for example from
non-resonant diagrams (e.g. $e^+e^-\to \mu^+\nu_\mu W^-$) and QCD contributions
$e^+e^-\to q \bar q gg,\; q \bar q q \bar q$ to the four-jet final state.
\end{itemize}
We will begin by studying (i) and (ii) in detail to identify the dominant
 effects, and later estimate the size of the background contributions.
Note that all the various theoretical components which we need for this study
already exist in the literature.

\subsection{The $W^+W^-$ off-shell cross section}
The leading-order cross section for off-shell $W^+W^-$
production was first presented in Ref.~\cite{MUTA}:
\begin{equation}
\sigma(s) = \int_0^s d s_1 \int_0^{(\sqrt{s}-\sqrt{s_1})^2} ds_2\;
\rho(s_1)\rho(s_2)\; \sigma_0(s,s_1,s_2) \; ,
\label{sig}
\end{equation}
where
\begin{equation}
\rho(s) = {1\over \pi} {\Gamma_W\over M_W}\; {s\over
(s-M_W^2)^2 + s^2 \Gamma_W^2/M_W^2 }  \; .
\label{rho}
\end{equation}
 Note that this expression  uses a $s$-dependent $W$ width,
\begin{equation}
\Gamma_W(s) = {s \over M_W^2} \; \Gamma_W
\; ,
\label{runningwidth}
\end{equation}
where $\Gamma_W \equiv \Gamma_W(M_W^2)$, as advocated in
Ref.~\cite{CAHN} for example. We will discuss this further below.

The cross section $\sigma_0$ can be written in terms of the $\nu$, $\gamma$ and
$Z$ exchange contributions and their interferences:
 \begin{equation}
\sigma_0(s,s_1,s_2) = {g^4\over 256 \pi s^2 s_1 s_2} \;
\left[ a_{\gamma\gamma} +  a_{ZZ} + a_{\gamma Z} +
a_{\nu\nu} +  a_{\nu Z} + a_{\nu\gamma} \right] \; ,
\label{siglo}
\end{equation}
where
$g^4=e^4/\sin^4\theta_W
%= 32 M_W^4 G_\mu^2
$. For completeness, the expressions for the various contributions
are listed in the Appendix.
The stable (on-shell) $W^+W^-$ cross section is simply
\begin{equation}
\sigma^{\rm on}(s) =  \sigma_0(s,M_W^2,M_W^2)\; .
\label{sigon}
\end{equation}
It will be an important factor in the discussion which follows
that near threshold the cross section is completely dominated
by the $t$-channel neutrino exchange diagram. This leads to an $S$-wave
threshold behaviour $\sigma_{t} \sim \beta$, whereas the $s$-channel
vector boson exchange diagrams give the characteristic
  $P$-wave behaviour $\sigma_{s} \sim \beta^3$.\footnote{We
  only consider unpolarized $e^+e^-$ collisions in this study.}
 This is illustrated in
Fig.~{\decomp}, which shows the  on-shell cross section
 decomposed in terms of the six contributions of Eq.~(\ref{siglo}).

\subsection{Higher-order electroweak corrections}

The complete set of $O(\alpha)$ next-to-leading order corrections to
$W^+W^-$ production has been calculated by several groups, for
the {\it on-shell} case only, see Ref.~\cite{BEENDENNER} and references
therein. There has been some progress \cite{AEPPLI1,AEPPLI2}
 with the off-shell (i.e. four
fermion production) corrections but the calculation is not yet complete.
However using the on-shell calculations as a guide, it is already possible to
predict some of the largest effects.
For example,
it has been shown that close to threshold the dominant contribution
comes from the Coulomb correction, i.e. the long-range electromagnetic
interaction between almost stationary heavy particles.
 Also important is the emission
of photons collinear with the initial state $e^\pm$
(`initial state radiation') which gives rise
to logarithmic corrections $\sim \alpha \ln(s/m_e^2)$. These leading logarithms
can be resummed to all orders, and incorporated for example
 using a `structure function'
formalism. In this case, the generalization from on-shell to off-shell
$W$'s appears to be straightforward. For the Coulomb corrections, however,
one has to be much more careful, since in this case the  inclusion of
 the finite $W$ decay width has a dramatic effect. Finally, one can
incorporate certain
 important higher-order fermion and boson loop corrections by a judicious
 choice of electroweak coupling constant. Each of these effects will be
discussed  in turn below.

In summary, certain $O(\alpha)$ corrections are already known
to be large, because their coefficients involve large factors
like $\log(s/m_e^2)$, $\sqrt{M_W/\Gamma_W}$, $m_t^2/M_W^2$,  etc.
Once these are taken into account, one may hope that the remaining
corrections are no larger than $\alpha \times $ constant.
When estimating the theoretical systematic uncertainty on the
$W$ mass in Section~\ref{conclusions}, we shall therefore assume an overall
uncertainty on the cross section
 of $\pm 1\%$ from the as yet uncalculated $O(\alpha)$ corrections.

\subsubsection{Coulomb corrections}

This section is essentially a summary of the results of
Refs.~\cite{FADIN1,FADIN3}, where the Coulomb corrections for on-shell
and off-shell $W^+W^-$ production have been discussed in detail,
and where a complete set of  references to earlier studies
can be found.

The result for on-shell $W^+W^-$ production is well known ---
the correction diverges as $1/v_0$ as the relative velocity $v_0$ of the
$W$ bosons approaches zero at threshold. Explicitly, to $O(\alpha)$,
\begin{equation}
\sigma_C(s) = \sigma_0(s)\;  [ 1+ \delta_C(s) ] \; ,
\label{coulon}
\end{equation}
where
\begin{equation}
\delta_C(s) = {\alpha\over v_0}\; \pi \; ,
\label{coulonbis}
\end{equation}
with the relative velocity given by
\begin{equation}
%v_0  = {4p_0\over \sqrt{s} } = 2\sqrt{1-{4 M_W^2\over s} }\; .
v_0  = 2\sqrt{1-{4 M_W^2\over s} }\; .
\end{equation}
As the threshold is approached from above, $\delta_C \to \infty$ and the
corrections  must be resummed to all orders, giving \cite{SOMMERFELD}
\begin{equation}
  1 +  {\alpha\over v_0} + \ldots \longrightarrow  {2 \alpha \pi / v_0 \over
  1  - \exp(-2 \alpha \pi/v_0)} \; .
\label{coulonall}
\end{equation}
Note that $\sigma_0 \sim v_0$ near threshold and so the
Coulomb-corrected cross section is formally
non-vanishing  when $\sqrt{s} = 2 M_W$.

For unstable $W^+W^-$ production the finite decay width $\Gamma_W$
 {\it screens} the Coulomb singularity \cite{FADIN1}, so that very close to
 threshold the perturbative expansion in $\alpha/v_0$ is effectively
 replaced  by an expansion in $\alpha \sqrt{M_W/\Gamma_W}$ \cite{FADIN3}.
 The net effect
 is a correction which reaches a maximum of approximately
 $+6\%$ in the threshold region   \cite{BARDIN1}.
Although this does not appear to be large, we will see below that it
changes  the threshold cross section by an amount equivalent
to a shift in $M_W$  of order $100$~MeV.
The result for off-shell $W^+W^-$ production is
(cf. Eqs.~(\ref{coulon},\ref{coulonbis}))
\begin{equation}
\sigma(s) = \int_0^s d s_1 \int_0^{(\sqrt{s}-\sqrt{s_1})^2} ds_2\;
\rho(s_1)\rho(s_2)\; \sigma_0(s,s_1,s_2) [ 1+ \delta_C(s,s_1,s_2)]
\label{sigcoul}
\end{equation}
where
\cite{FADIN3}\footnote{Note that in Ref.~\cite{BARDIN1}   a slightly
different version  of Eq.~(\ref{omega}) is derived, but the numerical
difference in the threshold region is very small. See Ref.~\cite{FADIN3}
for a critical discussion.}
\begin{equation}
\delta_C(s,s_1,s_2) = {\alpha\over v}\; \left[
\pi -2\; \mbox{arctan}\left({\vert\kappa\vert^2 -p^2  \over
 2 p\; {\rm Re}(\kappa ) }  \right) \right]
\label{omega}
\end{equation}
\begin{equation}
v = {4p\over \sqrt{s}}=2\sqrt{1-{(2s(s_1+s_2)-(s_1-s_2)^2)\over s^2} }
\end{equation}
\begin{equation}
\kappa = \sqrt{-M_W(E+i \Gamma_W)},\qquad
E = {s - 4M_W^2 \over 4 M_W}
\end{equation}

In Ref.~\cite{FADIN3} the above $O(\alpha)$
 result is generalized to all orders.
However the contributions from second order and above change
the cross section by $ \ll 1\%$ in the threshold region
\cite{KMS} and will be omitted in what follows.

\subsubsection{Initial state radiation}

Another  important class of electroweak radiative corrections
comes from the emission of photons from the incoming $e^+$ and $e^-$.
In particular, the emission of virtual and soft real photons with
energy $E < \omega$ gives rise to doubly logarithmic contributions
$\sim \alpha \ln(s/m_e^2) \ln(s/\omega^2)$ at each order in
perturbation  theory.  The infra-red ($\ln\omega$) logarithms cancel
when hard photon contributions are added, and the remaining
collinear ($\ln(s/m_e^2)$) logarithms can be resummed and incorporated in
the cross section using a `structure function'  \cite{FADIN4} (see
also Refs.~\cite{BERENDS1,CACCIARI,BERENDS2})
\begin{equation}
\sigma_{\mbox{\tiny ISR}} = \int_{s_{\rm min}}^s\ {ds'\over s}\;
F(x,s)\vert_{x=1-s'/s}\; \sigma(s')
\end{equation}
where \cite{BERENDS1}
\begin{equation}
F(x,s) = t x^{t-1}\; {\cal S} \; +\; {\cal H} \; ,
\label{factorisr}
\end{equation}
\begin{eqnarray}
{\cal S} & = &   1 +{3\over 4}\; t \; + \;  \left(
{9\over 32} -  {\pi^2 \over 12} \right)\; t^2  + {\alpha\over\pi}
\left({\pi^2\over 3} - {1\over 2}  \right) \; +\;  \ldots
\nonumber \\
{\cal H} & = &  \left( {x\over 2} -1 \right)\; t
\; + \;  \left[ \left({x\over 2}-1 \right)\ln x
- {1 +3 (1-x)^2\over 8x} \ln(1-x)   +{x\over 8} -{3\over 4} \right]\; t^2
\; +\;  \ldots  \; , \nonumber \\
\end{eqnarray}
with
\begin{equation}
t = {2\alpha \over \pi }  \left[ \ln \left({s \over m_e^2}\right)
-1  \right] \; .
\end{equation}
The ${\cal S}$ term comes from soft and virtual photon
emission, while the ${\cal H}$ term comes from hard collinear emission.
Contributions up to $O(t^2)$ are included, as well as a finite
soft-gluon contribution at $O(\alpha)$.

The incorporation of leading logarithmic effects in this way has been
widely discussed in the literature, see for example Ref.~\cite{BEENDENNER}
and references therein. In fact the form given above --- in which
the cross section is factorized into a structure function multiplying
the Born  cross section ---  is strictly
only valid for processes where there is no transfer of electric charge
 between the initial and final state particles, e.g. $e^+e^-
\to Z \to f \bar f$. Only the double logarithmic terms are in fact
`universal', i.e. process independent.  In general, the subleading
$O(\alpha)$ corrections do not factorize in the manner of
Eq.~(\ref{factorisr}). One possible refinement applicable
to $e^+e^- \to W^+W^-$  involves the `current
splitting technique' of Ref.~\cite{BARDIN2}, in which a charge
flow is artificially introduced into the $t$-channel neutrino
exchange in order that a gauge invariant initial-state radiation
contribution can be defined and factored out. In fact, close to
the $W^+W^-$ threshold the corrections obtained using
the current splitting and universal ISR
methods differ only at the $1\%$ level \cite{BARDIN2}. This difference is
in any case of  the same  order as  the neglected final state radiation
contributions.

\subsubsection{Improved Born approximation}

In the Standard Model, three parameters are sufficient to parametrize
the electroweak interactions, and the conventional choice is
$\{\alpha,G_\mu,M_Z\}$ since these are the three which
 are measured most accurately.
In this case the value of $M_W$ is a {\it prediction} of the model.
Radiative corrections to the expression
for $M_W$ in terms of these parameters
 introduce non-trivial dependences on $m_t$ and $M_H$, and
so a measurement of $M_W$ provides a constraint on these masses.
However the choice  $\{\alpha,G_\mu,M_Z\}$ does not appear to be
well suited to $W^+W^-$ production. The reason is that a variation
of the parameter $M_W$, which appears explicitly in the phase space
and in the matrix element,
has to be accompanied by an adjustment of the charged and neutral
weak couplings. Beyond leading order this is a complicated procedure.

It has been argued \cite{JEGERLEHNER} that a more appropriate choice
of parameters for LEP2 is the set $\{M_W,G_\mu,M_Z\}$
(the so-called $G_\mu$--scheme), since in this case
the quantity of prime interest is one
of the parameters of the model. Using the tree-level relation
\begin{equation}
g^2 = e^2/\sin^2\theta_W = 4 \sqrt{2} G_\mu  M_W^2
\label{gmuscheme}
\end{equation}
we see that the dominant $t$-channel neutrino
exchange amplitude, and hence the corresponding
contribution to the cross section,
depends only on the parameters $M_W$ and $G_\mu$.
It has also been shown \cite{JEGERLEHNER} that  in the
$G_\mu$--scheme there are no large next-to-leading order contributions to
the cross section which depend on $m_t$, either quadratically
or logarithmically. One can go further and choose the couplings
which appear in the other terms in the Born cross section
such that all  large corrections at
next-to-leading order are absorbed, see for example Ref.~\cite{DITTMAIER}.
However in this study we are only interested in the threshold
cross section, and so  we will simply use combinations of
 $e^2$ and $g^2$ defined  by Eq.~(\ref{gmuscheme})
for the neutral and charged weak
couplings which appear in the Born cross section, Eq.~(\ref{siglo}).

In summary, the most model-independent approach when defining
the parameters for computing the $e^+e^-\to W^+W^-$ cross section
appears to be the $G_\mu$--scheme, in which $M_W$ appears explicitly
as a parameter of the model. Although this makes a non-negligible
 difference when calculating the Born cross section, compared to using
$\alpha$ and $\sin^2\theta_W$
to define the weak couplings (see Table~\ref{tab:decomp} below),
a full next-to-leading-order
calculation will remove much of this scheme dependence \cite{JEGERLEHNER}.

\begin{table}[htb]
\centering
\begin{tabular}{|c|c|}
\hline
\rule[-1.2ex]{0mm}{4ex} parameter &  value \\
\hline
\rule[-1.2ex]{0mm}{4ex} $M_Z$ &  91.1888 \\
\rule[-1.2ex]{0mm}{4ex} $M_W$ &  80.23 \\
\rule[-1.2ex]{0mm}{4ex} $\Gamma_Z$ & 2.4974  \\
\rule[-1.2ex]{0mm}{4ex} $\Gamma_W$ &  2.078 \\
\rule[-1.2ex]{0mm}{4ex} $\alpha^{-1}$ &  137.0359895 \\
\rule[-1.2ex]{0mm}{4ex} $G_\mu$ & $1.16639 \times 10^{-5}$~GeV$^{-2}$  \\
\rule[-1.2ex]{0mm}{4ex}
$\sin^2\theta_W \equiv \sin^2\theta_W^{(\ell)\rm eff}$ &  0.2320 \\
\rule[-1.2ex]{0mm}{4ex} $m_e$ & $5.1099906\times 10^{-4}$ \\
\rule[-1.2ex]{0mm}{4ex} $(\hbar c)^2$ & $3.8937966 \times 10^{8}$~pb~GeV$^{2}$
\\
\hline
\end{tabular}
\caption{
Parameter values used in the numerical calculations. Masses and widths
are given in GeV.}
\end{table}

\subsection{Numerical evaluation of the cross section}

Figure~{\sigmaww}  shows the $e^+e^-\to W^+W^-$ cross section at LEP2 energies.
The different curves correspond to the sequential inclusion of the different
effects discussed in the previous section. The parameters used in
the calculation are listed in Table~1.
Note that both the initial state radiation and the finite width smear
the sharp threshold behaviour at $\sqrt{s} = 2 M_W$ of the on-shell
cross section.
 The different contributions
are quantified in Table~2, which gives the values of the cross section
in different approximations just above threshold ($\sqrt{s} = 161$~GeV)
and at the standard LEP2 energy of $\sqrt{s} = 175$~GeV.
At threshold we see that the effects of the intitial state radiation
and the finite width are large and comparable in magnitude.

\begin{table}[htb]
\centering
\begin{tabular}{|l|c|c|}
\hline
\rule[-1.2ex]{0mm}{4ex} $\sigma_{WW}$ & $\sqrt{s} = 161$~GeV &
 $\sqrt{s} = 175$~GeV \\
\hline
\rule[-1.2ex]{0mm}{4ex} $\sigma_0$ (on-shell, $\alpha$) & 3.813  & 15.092   \\
\rule[-1.2ex]{0mm}{4ex} $\sigma_0$ (on-shell, $G_\mu$) & 4.402  & 17.425   \\
\rule[-1.2ex]{0mm}{4ex} $\sigma_0$ (off-shell with
$\Gamma_W(M_W^2)$, $G_\mu$) & 4.747  & 15.873   \\
\rule[-1.2ex]{0mm}{4ex} $\sigma_0$ (off-shell with
$\Gamma_W(s)$, $G_\mu$) & 4.823  &   15.882 \\
\rule[-1.2ex]{0mm}{4ex} $\ldots\ +$ $O(\alpha)$ Coulomb & 5.122  & 16.311   \\
\rule[-1.2ex]{0mm}{4ex} $\ldots\ +$ $O(\alpha)$ Coulomb $+$ ISR&3.666&13.620\\
\hline
\end{tabular}
\caption{
Decomposition of the theoretical $e^+e^-\to W^+W^-$ cross section (in
picobarns)
as defined and discussed in the text, at two LEP2 collider energies.}
\label{tab:decomp}
\end{table}

Our primary aim is to investigate the dependence of the cross section
on $M_W$.  We do this by  taking the solid curve in Fig.~{\sigmaww}
($G_\mu$, finite width, initial state radiation, Coulomb correction)
as our `baseline' prediction.
Figure~{\ratios} shows  the cross sections for
$M_W = 80.0$ and $80.4$~GeV, normalized
to that for $M_W = 80.2$~GeV, as a function of the $e^+e^-$
collider energy $\sqrt{s}$.
As expected, the sensitivity to $M_W$ is greatest near threshold. Note that
for the on-shell cross section, with no initial state radiation,
the lower curve would decrease to $0$ at $\sqrt{s} = 160.8$~GeV
and the upper curve
would increase to $\infty$ at $\sqrt{s} = 160.4$~GeV. Figure~{\ratios} also
shows the
statistical uncertainty on the cross section
 ($\sim 1/\sqrt{N_{\rm ev}}$) which would be expected for
 $\lumi = 100\ \invpb$ at each of four separate collider
energies. The fact that this uncertainty increases as $\sqrt{s}$ decreases
is simply a reflection of the decreasing cross section. However, it is
immediately evident that the threshold region offers the highest statistical
sensitivity to $M_W$. The point of optimal sensitivity will
be discussed in the following section.

\section{Sensitivity of the cross section to $M_W$}

In this section we investigate  which LEP2 collider energy
offers the maximum precision on $M_W$ from a cross-section measurement.
There are several factors to take into consideration. The first
concerns the statistical uncertainty. Far above threshold
the number of $W^+W^-$ events is large but this is outweighed by
the very weak $M_W$ dependence. Far below the (nominal) threshold
there are too few events to achieve any precision. The second concerns
the systematic uncertainties. The important non-$W^+W^-$ background processes
do not have the threshold energy dependence of the signal, and therefore
constitute an increasingly large fraction of the event sample as the threshold
 energy is approached from above. The optimal collider energy must therefore
 take all these competing factors into account.

\subsection{Statistical uncertainty}
If $\lumi$ is available for measuring the cross section with an
efficiency $\epsilon$ at a collision energy $\sqrt{s}$, then
the statistical error on $M_W$ can be estimated as
\begin{equation}
\Delta M_{\rm stat} = \left\vert {d \sigma \over d M}
\right\vert^{-1}\; \Delta\sigma
 =   \left\vert {d \sigma \over d M}\right\vert^{-1}\;
\sqrt{{\sigma\over \epsilon \cdot \lumi }} \equiv
K \left[ \epsilon \cdot \lumi  \right]^{-{1\over 2}}\; ,
\end{equation}
where
\begin{equation}
K  = \sqrt{\sigma}\; \left\vert {d \sigma \over d M}
\right\vert^{-1}
\end{equation}
is a measure of the sensitivity to $M_W$.
The dependence of $K$ on $\sqrt{s}$ is shown in Fig.~{\statsens}.
There is evidently  a fairly sharp minimum just
($\sim 500$~MeV) above the nominal
threshold of $\sqrt{s} = 2M_W$, at which point
\begin{equation}
\Delta M_{\rm stat}\  =\  90\; {\rm MeV}\
\left[ {\epsilon \cdot  \lumi \over
100\; {\rm pb}^{-1} }  \right]^{-{1\over 2}} \; .
\label{eq:stat}
\end{equation}

\subsection{Systematic uncertainties}
If the cross section is subject to an overall {\it multiplicative}
factor $C$ with an uncertainty $\Delta C$
(high-order corrections, luminosity, efficiency, ...)
then the associated systematic error is
\begin{equation}
\Delta M_{\rm sys}\ =\ \left\vert {d \sigma \over d M}
\right\vert^{-1}\; \sigma \ {\Delta C \over C}\ \equiv
\ J\ {\Delta C \over C} \; ,
\end{equation}
where
\begin{equation}
J = \sigma  \left\vert {d \sigma \over d M}
\right\vert^{-1}\; .
\end{equation}
The quantity $J$ is shown in Fig.~{\multsens} as a function
of $\sqrt{s}$. Again we see that the curve has a minimum in
the threshold region. At the energy ($\sqrt{s} \simeq 161$~GeV)
at which  $K$ is minimized we see that $J \simeq 1.7$~GeV, and so
\begin{equation}
\Delta M_{\rm sys}\ \simeq\ 17\; {\rm MeV}
\ \left[ {\Delta C \over C} \times 100\% \right] \; .
\end{equation}
For an {\it additive} uncertainty (for example from the subtraction
of a non-$WW$ background cross section, see below) the contribution
to the systematic error is simply
\begin{equation}
\Delta M_{\rm sys}\ =\ \left\vert {d \sigma \over d M}
\right\vert^{-1}\; \Delta\sigma \ \equiv \ L \; \Delta\sigma \; .
\end{equation}
The quantity  $L$ is shown in Fig.~{\syssens} as a function of $\sqrt{s}$
in the threshold region. The curve has a similar shape to the
statistical sensitivity function shown in Fig.~{\statsens},
although the minimum is not quite as sharp.
At the energy  ($\sqrt{s} \simeq 161$~GeV) at which $K$ is minimized
we see that
\begin{equation}
\Delta M_{\rm sys}\
 = \ 470\; {\rm MeV} \ \left[{\Delta \sigma \over 1\ {\rm pb}}
  \right] \; .
\label{sys}
\end{equation}

Finally, there is in principle a systematic uncertainty in $M_W$
arising from an uncertainty  in the beam energy.
It is already apparent  from Fig.~{\fivemasses} that a shift in the beam energy
is equivalent to a shift in $M_W$. Quantitatively,
\begin{equation}
\Delta M_{\rm sys}\ =\ \left\vert {d \sigma / d E_{\rm beam} \over
d \sigma / d M}
\right\vert\  \Delta E_{\rm beam} \; .
\label{beam}
\end{equation}
The ratio of derivatives is almost independent of energy in the threshold
region and has a numerical value very close to 1, i.e.
\begin{equation}
\Delta M_{\rm sys}\
  \simeq\ 1.0 \ \Delta E_{\rm beam} \; .
\label{beamerror}
\end{equation}
Current estimates suggest that
 a $\Delta E_{\rm beam}$ of less than
15~MeV can be achieved \cite{MYERS}, and so
the systematic error from this source
is not expected to be particularly significant.

\subsection{Dependence of the cross section on $\Gamma_W$}

In the Standard Model, the decay width of the $W$ is
\begin{eqnarray}
\Gamma_W & = & [3 + 6\{ 1 + \alpha_s(M_W)/\pi + ... \}] \ \Gamma (W \to \ell
\nu)        \nonumber \\
 & = & [0.1084 \pm 0.0002 ] \ {G_{\mu} M_W^3 \over \sqrt{2}\;  6 \pi}\;
\left(1 + \delta^{\rm SM}\right) \; ,
\end{eqnarray}
where $\alpha_s(M_W) = 0.12 \pm 0.01$ is used in the
leptonic branching ratio. The higher order electroweak corrections
depend weakly on $m_t$ and $M_H$, and for canonical values of these
$\delta^{\rm SM} = -0.0035$ \cite{DENNER,SACK} (see also \cite{ROSNER}).
 The main uncertainty
therefore comes from the value of $M_W$ itself,
\begin{equation}
\Gamma_W  =  2.078 \left({M_W \over 80.23\; {\rm GeV}}\right)^3\ {\rm
GeV}\; .
\label{gammawsm}
\end{equation}
If we use the current world average value
 (Eq.~(\ref{currentmw})) for $M_W$ then
$\Gamma_W = 2.078 \pm 0.014$~GeV. In contrast, the existing measurements
of the $W$ width from the CDF collaboration
at the Tevatron $p \bar p$ collider are much less precise:
\begin{equation}
\Gamma_W =  2.064 \pm 0.061 ({\rm stat}) \pm 0.059 ({\rm sys})\ {\rm GeV}
\end{equation}
{}from the ratio of $W$ and $Z$ cross sections method \cite{CDFGW1}, and
\begin{equation}
\Gamma_W =  2.11 \pm 0.28 ({\rm stat}) \pm 0.16  ({\rm
sys})\ {\rm GeV}
\end{equation}
{}from the shape of the lepton $p_T$ distribution near the end-point
\cite{CDFGW2}. Both measurements
are evidently in good agreement with the Standard Model
prediction.  The weighted average of the two measurements is
$\Gamma_W = 2.067 \pm 0.082$~GeV.

Concerning the impact of $\Gamma_W$  on the determination
of $M_W$, there are two points of view that one can adopt. In the
context of a Standard Model fit, one can simply use the
result Eq.~(\ref{gammawsm}) when studying the variation of the
cross section with $M_W$ near threshold. In this case, the effect
on the cross section as $\Gamma_W$ varies with $M_W$ is {\it much} smaller
than the variation with $M_W$ directly, so that for all practical
purposes $\Gamma_W$ can be fixed at its nominal value of $2.078$~GeV.

A less model dependent approach is to study the variation of the cross
section when $\Gamma_W$ is varied in its allowed experimental range,
independently of $M_W$.
Figure~{\widthvar} shows the cross section near threshold
 for $\Gamma_W = 2.078 \pm 0.1$~GeV, as a function of $\sqrt{s}$
for fixed $M_W = 80.23$~GeV. Note that there is a  stable point
at $\sqrt{s} \sim 162$~GeV where the cross section is approximately
independent of $\Gamma_W$. From Fig.~{\widthvar} we can
derive\footnote{Note that below $\sqrt{s} \simeq 162$~GeV, {\it increasing} the
width has the same effect on the cross section as {\it decreasing} the mass.}
\begin{equation}
\Delta M \  = \  \left\vert {d \sigma / d \Gamma \over d \sigma / d M}
\right\vert \;
\Delta \Gamma \ \simeq \ 0.16\; \Delta\Gamma \; ,
\end{equation}
where the derivatives are evaluated at $\sqrt{s} = 161$~GeV.
Hence the uncertainty on $M_W$ from the {\it current} measurement
of $\Gamma_W$ is $\pm 13$~MeV. It is expected, however, that the
precision of the $p \bar{p}$ collider $\Gamma_W$
determination  will improve
significantly in the next few years \cite{SACHA}.

A final comment concerns the use of the {\it running} $W$ width
$\Gamma_W(s)$  (Eq.~(\ref{runningwidth})) in our calculations.
It is clear from Table~\ref{tab:decomp}, where the cross section
is computed for both fixed and running widths, that this {\it does} make
a non-negligible difference in the threshold region.
However the effect can be largely understood as a shift
in the $W$ mass, defined as the position of the pole in the propagator:
\begin{equation}
(s-M^2)^2 + (s\Gamma/M)^2 = (1+\Gamma^2/M^2)\; (s-\overline{M}^2)^2
+ (\overline{M}\Gamma)^2
\end{equation}
where $\overline{M} = M(1+\Gamma^2/M^2)^{-1/2} \approx M -
(\Gamma^2/2M) + \ldots \approx M - 27$~MeV. In fact, this is simply
a question of the {\it definition} of the $W$ mass, and is not really
an additional uncertainty. The conventional procedure (used for example
in the $Z$ mass measurement at LEP) is to adopt the `running width'
definition.

\subsection{Background contributions}

Any additional background contributions to the cross section
 will reduce
the sensitivity  to $M_W$. There are two main
effects: (i)  the statistical error on $\sigma_{WW}$ is increased by
 additional non-$W^+W^-$ contributions to the event sample, and
(ii) an uncertainty in the subtracted background cross section
becomes an uncertainty on $\sigma_{WW}$ and hence on $M_W$.
Let us assume that the background cross section $\sigma_{\rm bkd}$ and
its uncertainty $\Delta\sigma_{\rm bkd}$ are small compared
to $\sigma_{WW}$. It is straightforward to show that the
statistical uncertainty on $M_W$ is increased to
\begin{equation}
\Delta M_{\rm stat} \longrightarrow \Delta M_{\rm stat}
\; \cdot \; \sqrt{1 + {\sigma_{\rm bkd} \over \epsilon\cdot
\sigma_{WW}}} \; ,
\label{bkdstat}
\end{equation}
where $\epsilon$ is the efficiency for detecting the $W^+W^-$ signal.
The contribution to the systematic error from the uncertainty
on the background cross section is given by Eq.~(\ref{sys}),
with $\Delta\sigma = \Delta\sigma_{\rm bkd}$.

A detailed quantitative assessment of all sources of background
events is beyond the scope of the present work. It is however possible
to give some estimates as to the likely size of the most important
contributions. We can begin by classifying backgrounds into
two types: {\it irreducible} and {\it reducible}.
By the  former we mean a four-fermion final state with
the same flavour structure as a  $W^+W^-$ decay.
This is dominated by  non-resonant electroweak contributions
to the $4f$ final state. For example, there is a non-resonant
contribution  to $e^+e^- \to W^+W^- \to e^+ \nu d \bar u$
{} from  $e^+e^- \to \gamma^*,Z^* \to e^+ \nu W^- \to e^+ \nu d \bar u$
where the $W^-$ is radiated off an outgoing $e^-$ produced
at the $\gamma/Z$ vertex.\footnote{This corresponds to a `reindeer'
diagram in the language of Ref.~\cite{BARDIN3}.}
Such non-resonant contributions have been studied in detail in
Refs.~\cite{BARDIN3} and \cite{EXCALIBUR}.
The actual size of the additional contributions
to the $WW$  cross section
depends on the particular fermion channels, the collider energy
and on possible final-state cuts.
A semi-analytical calculation of the
background cross section \cite{BARDIN3} for a range
of $4f$ final states (non-identical $f$, no $e$
or $\nu_e$) shows that in the threshold region $\sigma_{\rm bkgd}(4f)
/\sigma_{WW} \ll 1\%$, and therefore the  effect on the $M_W$
determination is negligible. It is possible that final states
such as $e- \nu_e W^+$ pose more of a problem, because of the
$t$-channel photon exchange contribution. In this case one needs
a careful choice of cuts to eliminate the potentially large
background cross section,
see for example Ref.~\cite{EXCALIBUR}.

Reducible backgrounds have a final state which does not
correspond to any $WW\to 4f$ decay channel but which in practice
cannot be distinguished. By far the most important of these appears to
be QCD four-jet production, $e^+e^- \to \gamma^*,Z^*\to
q\bar q gg, q \bar q q \bar q$, which is a background
to $e^+e^- \to W^+W^-\to  q \bar q' q \bar q'$. Since  $q \bar q gg$
production dominates the QCD cross section,
there is practically no interference between the QCD and $W^+W^-$
final states. Although the QCD four-jet background does not appear
to cause problems in reconstructing $M_W$ from four-jet
final states well above threshold (see for example Ref.~\cite{DELPHI}),
it is potentially more serious for the threshold measurement,
since the $W^+W^-$ cross section is much smaller at threshold
 and the QCD cross section increases with decreasing $\sqrt{s}$.

In order to gauge the size of the QCD background contribution,
we have calculated\footnote{We use the matrix elements from
Ref.~\cite{CALKUL}.} the cross sections for $e^+e^- \to
q\bar q gg, q \bar q q \bar q$ in the threshold region.
To define a finite-cross section we must introduce
cuts which keep the massless quarks and gluons energetic and non-collinear.
We do this with a simple JADE-type algorithm \cite{JADE}, i.e. we
require that $(p_i+p_j)^2 > y_{\rm cut} s $, where $i,j = q,g$,
and $y_{\rm cut}$ is a dimensionless parameter.
Since this is a tree-level calculation at the parton level,
it is  not  straightforward to  relate
 it to the measured jet cross section. However this uncertainty
 can be significantly reduced by
normalizing the parton-level prediction  to the  four-jet
production rate already measured at LEP. Figure~{\lepfourj} shows the
fraction of four-jet events measured by OPAL \cite{OPAL} at $\sqrt{s} = M_Z$
as a function of $y_{\rm cut}$. The curves are the theoretical
predictions for two different values of the argument of the
strong coupling $\alpha_s$ ($\mu=\sqrt{s},\; \sqrt{s}/2$) chosen
to give a reasonable spread of agreement with the data.
Figure~{\lepiifourj} shows the corresponding (absolute) four-jet
cross sections at $\sqrt{s} = 161$~GeV. In addition to the jet
separation cut, we have required that in each event there are two
jet pairs each with an invariant mass within 10~GeV of $M_W$. The lower
dashed lines are the QCD predictions corresponding to the
`fits' in Fig.~{\lepfourj}. The dotted line corresponds to $W^+W^- \to
4$~jets with no cuts, i.e. the total cross section multiplied
by a branching ratio of 0.46 for the $q \bar q q \bar q $
final state. The solid line is the $W^+W^-$ prediction with both types
of cut included. We see that the QCD curves fall more rapidly with increasing
$y_{\rm cut}$ than the $W^+W^-$ curve, as expected from the
different singularity
structure of the corresponding matrix elements.

Taking the average of the dashed curves as our estimate of the QCD
background, we deduce from Fig.~{\lepfourj} that the square root
factor in Eq.~(\ref{bkdstat}) falls slowly  from about 1.04 at $y_{\rm cut} =
0.01$
to about 1.01 at $y_{\rm cut} = 0.07$. However this improvement
at large $y_{\rm cut}$ is more than offset by the overall decrease
in the efficiency for detecting the signal, which is 0.93 at
$y_{\rm cut} = 0.01$ but only 0.20 at $y_{\rm cut} = 0.07$.
We conclude that a small $y_{\rm cut}$ value is preferred. If we
take $y_{\rm cut} = 0.01$ as representative, we obtain
\begin{eqnarray}
\sigma_{WW}(4j,\; \mbox{no cuts}) & = & 2.20\; {\rm pb} \nonumber \\
\sigma_{WW}(4j) & = & 1.90\; {\rm pb} \quad (\epsilon = 0.87) \nonumber \\
\sigma_{QCD}(4j) & = & 0.15 \pm 0.03 \; {\rm pb} \; .
\label{backsummary}
\end{eqnarray}

To estimate the overall efficiency  we must  include
also  the $q \bar q \ell (=e, \mu) \nu$ final state. Let us assume
for simplicity that the efficiency for this channel is $87\%$ also.
Then the {\it total} efficiency can be estimated as\footnote{
Note that if electrons and muons from $W\to \tau \nu$ decay are also
included, this number will increase slightly.}
\begin{equation}
\epsilon_{\rm all}  = 0.87 \times [\; BR(W^+W^-\to q \bar q q \bar q)
\; + \; BR(W^+W^-\to q \bar q \ell \nu) \; ] = 0.65 \; .
\end{equation}
Note that a 65\% efficiency leads, via Eq.~(\ref{eq:stat}), to an increase
in the statistical error by a factor  1.24.
If we assume that the background is indeed dominantly due to QCD four-jet
production, then there is a further increase of 3\% from the
dilution factor in Eq.~(\ref{bkdstat}).

\section{A strategy for scanning}

It is anticipated that a total luminosity of $O(500)$~pb$^{-1}$
 will be obtained at LEP2.
There are strong physics arguments (Higgs and other new particle
searches, anomalous $WWV$ coupling  limits, \ldots )
for running at the highest possible
energies,  $\sqrt{s} \sim 175 - 200$~GeV, rather than at threshold.
However, given the importance of a precision $M_W$ measurement and
remembering that there {\it may} be a difficulty in controlling
the  QCD `reconnection effects' for the $q \bar q q \bar q$ final state,
it is worth giving serious consideration to using some of the available
luminosity in the threshold region. Realistically,
$\lumi = 50 - 100$~pb$^{-1}$ could be regarded  as reasonable, in which case
the uncertainty on $M_W$ obtained from the threshold cross section
is predominantly statistical.
One therefore has to think carefully about which collider energy (energies)
are optimal. It is clear from Figs.~{\ratios} and {\statsens} that the only
sensible strategy is to run just above the nominal threshold.
Notice from Fig.~{\fivemasses} that with this amount of luminosity
there is no question of obtaining any useful
information on $M_W$ from the {\it shape} of the cross section alone, and so
running at a  single collider  energy, or narrow range of energies,
is sufficient. In doing so we are relying on an absolute theoretical
prediction and an absolute cross-section measurement.

Of course the value of the collider energy which  corresponds to
 maximum statistical sensitivity depends on
$M_W$ itself. {\it However, we already knows the $W$ mass well enough
to be able to specify this energy in advance.} The sensitivity parameter
$K$ of Fig.~{\statsens} has a formal minimum at about 500~MeV above the
nominal threshold of $\sqrt{s}= 2M_W$,
but the variation with $\sqrt{s}$ is only slight in a region
$ \pm O(500)$~GeV on either side of this.
Given that when LEP2 starts operation
the error on $M_W$ from the $p \bar p $ collider could be smaller by a factor
of two than its present value of $\pm 180$~MeV, one will know the optimal
collider energy accurately enough. In addition, an initial LEP2
run at a much higher
energy will allow $M_W$
to be determined  even  more precisely from  reconstructing the final state.

The situation would be very different if one had no {\it a priori}
knowledge of $M_W$. Then it would be necessary
 to devise a method of locating
the collider energy of maximum sensitivity without wasting luminosity.
This point has been addressed in detail in Ref.~\cite{CHEN}. It turns
out that a `data driven scanning strategy' (the BES method) can be devised.
According to this one makes a sequence of measurements at different,
carefully specified collider energies,  continually
updating the $M_W$ determination, until one  eventually
arrives at the optimal energy point
just above the nominal threshold. The technique has been used
successfully to measure $m_\tau$ at lower energies \cite{BES}.

\section{Conclusions}
\label{conclusions}

In this paper we have investigated in detail the measurement
of $M_W$ from the $W^+W^-$ cross section near threshold
at LEP2.  We have estimated the statistical and systematic uncertainties
and shown that it will be the former which will dominate the error.
It is therefore important to choose the collider energy carefully to
maximise the sensitivity. We have shown that the optimal point
is approximately 500~MeV above the nominal threshold
energy, $\sqrt{s} = 2 M_W$. Based on current $M_W$ measurements,
this corresponds to $\sqrt{s} \approx 161$~GeV. The actual value is
not crucial within $\pm 500$~MeV or so. Given that the existing
experimental error on $M_W$ is well within this range, there does not
appear to be any need to `scan' the threshold  region to locate
the most sensitive point. Our estimate for the statistical uncertainty
on $M_W$ at $\sqrt{s}  =  161$~GeV
is given in Eq.~(\ref{eq:stat}). If we assume
\begin{itemize}
\item[{(i)}] four experiments each obtaining $\lumi$ of integrated
luminosity,
\item[{(ii)}] an overall efficiency of 65\%, corresponding to detecting
the $q \bar q q \bar q $ and $q \bar q \ell (=e,\mu) \nu$ final states
with an efficiency of 87\%,
\end{itemize}
then our estimate is
\begin{equation}
\Delta M_{\rm stat}\  =\  56\; {\rm MeV}\
\left[ {\lumi \over
100\; {\rm pb}^{-1} }  \right]^{-{1\over 2}} \; .
\label{eq:statagain}
\end{equation}
Depending on the size of the background cross section --- see below ---
this number may increase by a few per cent.

\begin{table}[htb]
\centering
\begin{tabular}{|l|c|}
\hline
\rule[-1.2ex]{0mm}{4ex} source &  $\Delta M_W$ \\
\hline
\rule[-1.2ex]{0mm}{4ex} HO corrections &
$\Delta\sigma / \sigma (\%)\; \times\; 17 $~MeV \\
\rule[-1.2ex]{0mm}{4ex} luminosity &
$\Delta{\cal L} / {\cal L}(\%)\; \times\; 17 $~MeV \\
\rule[-1.2ex]{0mm}{4ex} efficiency &
$\Delta\epsilon / \epsilon(\%)\; \times\; 17 $~MeV \\
\rule[-1.2ex]{0mm}{4ex} background &
$\Delta\sigma \; \times\; 470 $~MeV/pb \\
\rule[-1.2ex]{0mm}{4ex} beam energy &
$1.0 \; \Delta E_{\rm beam}$ \\
\rule[-1.2ex]{0mm}{4ex} decay width &
$ 0.16\; \Delta \Gamma_W$ \\
\hline
\end{tabular}
\caption{
Summary of systematic uncertainties on $M_W$ from a measurement
of the $W^+W^-$ cross section at
$\protect{\sqrt{s}}\; \protect{\simeq}\; 161$~GeV.}
\label{tab:sysconc}
\end{table}

We have also investigated  various sources of systematic uncertainty,
and our results are summarized in Table~\ref{tab:sysconc}.
{\it There does not appear to be any single dominant source.}
If we assume
\begin{itemize}
\item[{(i)}] by the time LEP2 comes into operation the theoretical
cross section will be known to 1\% accuracy, which
is hopefully conservative,
\item[{(ii)}] an uncertainty of 1\% on both the efficiency
and the luminosity,
\item[{(iii)}] an uncertainty on the background cross section
of $\Delta\sigma_{\rm bkd} = \pm 0.03$~pb, see Eq.~(\ref{backsummary}),
\item[{(iv)}] a beam energy uncertainty of 15~MeV, which again is
probably conservative,
\item[{(iv)}] either the $W$ width is calculated in the Standard Model
or will be measured with equivalent precision before LEP2,
\end{itemize}
then from Table~\ref{tab:sysconc} our estimate for the overall systematic
uncertainty  is\footnote{combining the individual contributions
in quadrature}
\begin{equation}
\Delta M_{\rm sys}\  =\  36\; {\rm MeV}\; .
\label{eq:sysagain}
\end{equation}

We cannot of course claim to have done a comprehensive study of
{\it all} of the above effects, statistical or systematic, theoretical
or experimental. However we have tried to use reasonable estimates
 to set benchmarks against which future, more complete studies can be
 compared. It should be clear from the previous sections where more
work needs to be concentrated. Our treatment of background contributions
has been superficial, to say the least.
Nevertheless, our overall conclusion  is that the idea of running
LEP2 at the $W^+W^-$ threshold in a dedicated attempt to measure
$M_W$ from the cross section is worthy of serious consideration.
With an overall luminosity of order 100~pb$^{-1}$ per experiment,
the target error on $M_W$ of $\pm 50$~MeV could well be achievable.

A final comment concerns the impact of new physics beyond
the Standard Model on this method of measuring the $W$ mass.
As already stated, we are relying on an absolute cross-section
calculation, and therefore any additional new-physics contributions
will distort the $M_W$ measurement. In particular, we have shown that
at $\sqrt{s} = 161$~GeV a small shift in the cross section by an amount
$\Delta\sigma_{\rm new}$ is equivalent to a shift in $M_W$ of
$ 0.47\; {\rm MeV\; fb}^{-1} \times \Delta\sigma_{\rm new}$.
However we have seen (Fig.~{\decomp}) that at threshold energies the
$t$-channel
neutrino-exchange diagram is completely dominant. The typical  new-physics
scenarios which have been considered would be expected to affect only
the $s$-channel contributions; for example, anomalous $\gamma W^+W^-$,
 $Z W^+W^-$ couplings or new heavy boson production $e^+e^-\to Z' \to 4f$.
The corresponding contributions to the cross section would therefore
be suppressed by at least one  additional power of $\beta$. In other words,
the threshold region is probably safe from most types of new-physics
contamination, at least at a level which would cause a significant
shift in $M_W$.
In any case, it is likely that a threshold energy LEP2 run would take
place {\it after}  an initial run at  higher energy, in which case
the presence  of new physics   would almost certainly
have already  been detected. Nevertheless, a more detailed study
of new-physics contributions in the threshold region would be very desirable
\cite{KUNSZT}.

\section*{\Large\bf Acknowledgements}

\noindent I am grateful to the UK PPARC
 for  financial support in the form of a Senior Fellowship.
Useful discussions with Alain Blondel, Robert Cahn, Stavros Katsanevas,
Valery Khoze, Zoltan Kunszt, Alan Martin and Andrea Valassi are acknowledged.
This work was supported in part by the EU Programme
``Human Capital and Mobility'', Network ``Physics at High Energy
Colliders'', contract CHRX-CT93-0537 (DG 12 COMA).
\goodbreak

\newpage
\setcounter{section}{1}
\setcounter{equation}{0}
\renewcommand{\thesection}{\Alph{section}}
\renewcommand{\theequation}{\Alph{section}\arabic{equation}}
\section*{Appendix}
The following expressions, when substituted into Eqs.~(\ref{sig}) and
(\ref{siglo}), give the leading-order off-shell $e^+e^-\to W^+W^-$ cross
section. The results are taken from Ref.~\cite{MUTA}.
\begin{eqnarray}
a_{\gamma\gamma} &=& 8 \sin^4\theta_W \; {1 \over s^2}
\; G_1(s,s_1,s_2)  \nonumber \\
a_{ZZ} &=& {1\over 2} (v_e^2+a_e^2)\;
{ 1 \over (s-M_Z^2)^2+ M_Z^2\Gamma_Z^2(s) } \; G_1(s,s_1,s_2)  \nonumber \\
a_{\gamma Z} &=& 4 v_e \sin^2\theta_W\; {1\over s} \;
{ s-M_z^2 \over (s-M_Z^2)^2+ M_Z^2\Gamma_Z^2(s) } \; G_1(s,s_1,s_2)  \nonumber
\\
a_{\nu\nu} &=&  \; G_2(s,s_1,s_2)  \nonumber \\
 a_{\nu Z} &=& -(v_e+a_e)\; { s-M_z^2 \over (s-M_Z^2)^2+ M_Z^2\Gamma_Z^2(s) }
 \; G_3(s,s_1,s_2)  \nonumber \\
a_{\nu\gamma} &=& -4 \sin^2\theta_W\; {1 \over s} \; G_3(s,s_1,s_2)
\end{eqnarray}
\begin{eqnarray}
G_1 &=& \lambda^{3/2} \left[ {\lambda\over 6}+
2(ss_1+ss_2+s_1s_2)\right]  \nonumber \\
G_2 &=& \lambda^{1/2} \left[ {\lambda\over 6}+
2(ss_1+ss_2-4s_1s_2) + 4s_1s_2(s-s_1-s_2) F \right]   \nonumber \\
G_3 &=& \lambda^{1/2} \left[ {\lambda\over 6}(s+11s_1+11s_2)
+2s(s_1^2+s_2^2+3s_1s_2)\right. \nonumber \\
& & \left.-2(s_1^3+s_2^3)
- 4s_1s_2(ss_1+ss_2+s_1s_2) F \right] \; ,
\end{eqnarray}
where $a_e = 1$, $v_e = 1-4 \sin^2\theta_W$ and
\begin{equation}
\lambda(s,s_1,s_2) = s^2 + s_1^2 + s_2^2 -2(s s_1 + s s_2 + s_1 s_2 )
\end{equation}
\begin{equation}
F(s,s_1,s_2) =  {1\over \sqrt{\lambda}}\; \ln \left(
{s-s_1-s_2 +\sqrt{\lambda}  \over s-s_1-s_2 - \sqrt{\lambda} } \right) \; .
\end{equation}

\newpage

\section*{Figure Captions}
\begin{itemize}

\item [{[\decomp]}] Decomposition of the Born $e^+e^-\to W^+W^-$
(on-shell) cross section into the contributions from $t$-channel
neutrino and $s$-channel $\gamma^*,Z$ exchange and their
interferences.

\item [{[\sigmaww]}]
The cross section for $e^+e^- \to W^+ W^-$
in various approximations: (i) Born (on-shell) cross section,
(ii) Born (off-shell) cross section, (iii)
with first-order Coulomb corrections, and (iv) with initial-state
radiation. The parameter values are given in the text.

\item [{[\ratios]}]
The cross sections (off-shell, with Coulomb and ISR
corrections) for $M_W = 80.0$ and $80.4$~GeV, normalized
to that for $M_W = 80.2$~GeV, as a function of the $e^+e^-$
collider energy $\sqrt{s}$. Also shown are `data points' with
statistical errors corresponding to $\lumi = 100\ \invpb$ at each
energy.

\item [{[\fivemasses]}] The cross sections (off-shell, with Coulomb and ISR
corrections) for $M_W = 78.8 +  0.2 n $~GeV ($0 \leq n\leq 4$)
as a function of the $e^+e^-$
collider energy $\sqrt{s}$ in the threshold region.

\item [{[\statsens]}] The statistical sensitivity factor $K =
\sqrt{\sigma}\; \left\vert {d \sigma / d M}\right\vert^{-1}$
as a function of the collider energy $\sqrt{s}$. The arrow indicates the
 position of the nominal threshold, $\sqrt{s} = 2 M_W$.

\item [{[\multsens]}] The multiplicative systematic sensitivity factor $J =
\sigma \;\left\vert {d \sigma / d M}\right\vert^{-1}$
as a function of the collider energy $\sqrt{s}$. The position of the nominal
threshold is indicated.

\item [{[\syssens]}] The additive systematic sensitivity factor $L =
\left\vert {d \sigma / d M}\right\vert^{-1}$
as a function of the collider energy $\sqrt{s}$. The position of the nominal
threshold is indicated.

\item [{[\widthvar]}] Dependence of the threshold cross section
on the width $\Gamma_W$.

\item [{[\lepfourj]}] The
fraction of four-jet events measured by the  OPAL
collaboration \cite{OPAL} at $\sqrt{s} = M_Z$
as a function of $y_{\rm cut}$. The curves are the theoretical
predictions using the $O(\alpha_s^2)$ $q\bar q q\bar q$ and $q \bar q g g$
matrix elements with two different values of the argument $\mu$ of the
strong coupling $\alpha_s$.

\item [{[\lepiifourj]}]
Four-jet cross sections from $W^+W^-$ and QCD four-parton production
at $\sqrt{s} = 161$~GeV, as a function of the jet separation
parameter $y_{\rm cut}$.
An additional cut requires that in each event there are two
jet pairs each with an invariant mass within 10~GeV of $M_W$. The lower
dashed lines are the QCD predictions corresponding to the
curves in Fig.~{\lepfourj}. The dotted line corresponds to $W^+W^- \to
4$~jets with no cuts,
and the solid line has both types of cut included.

\end{itemize}

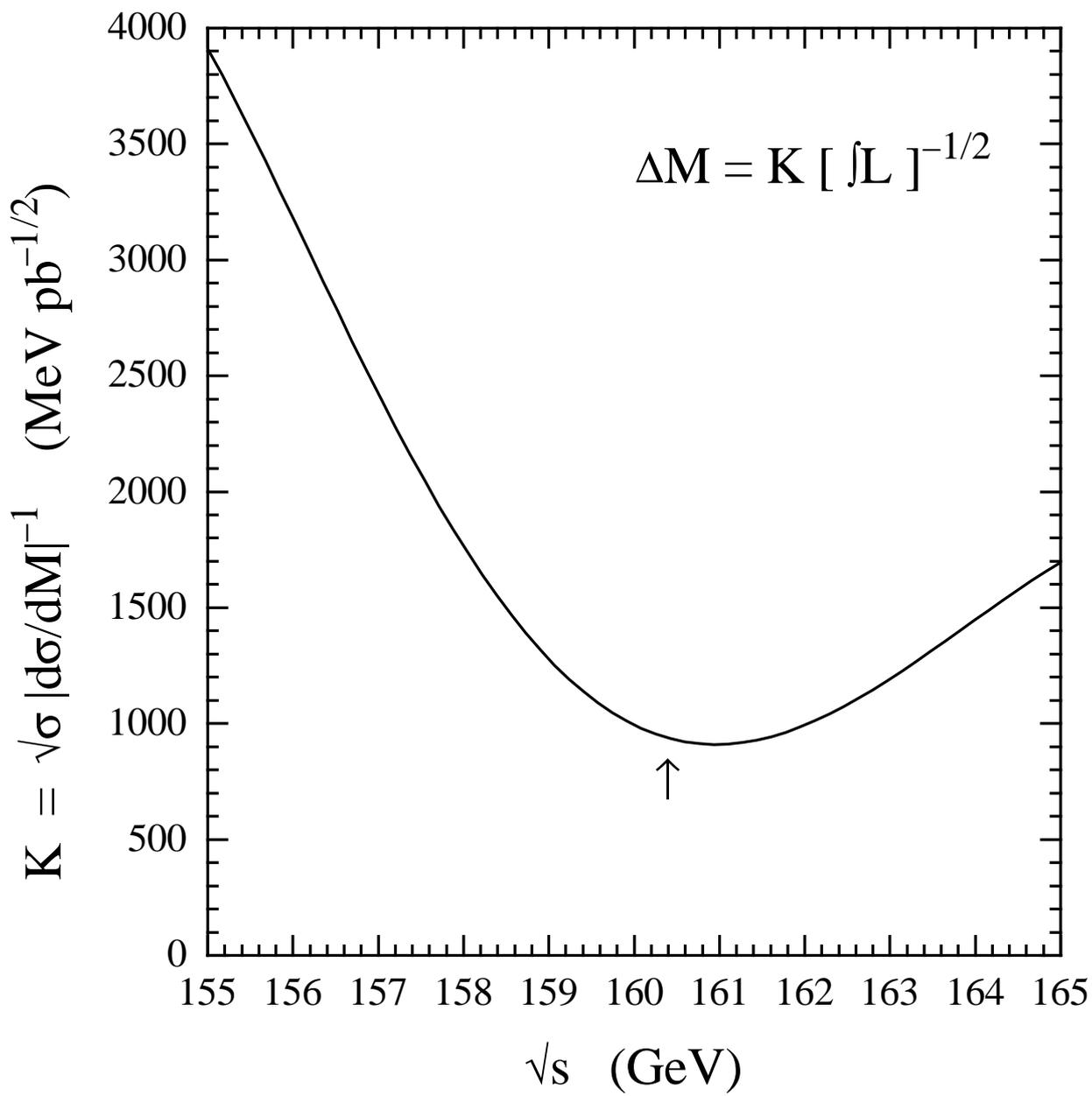
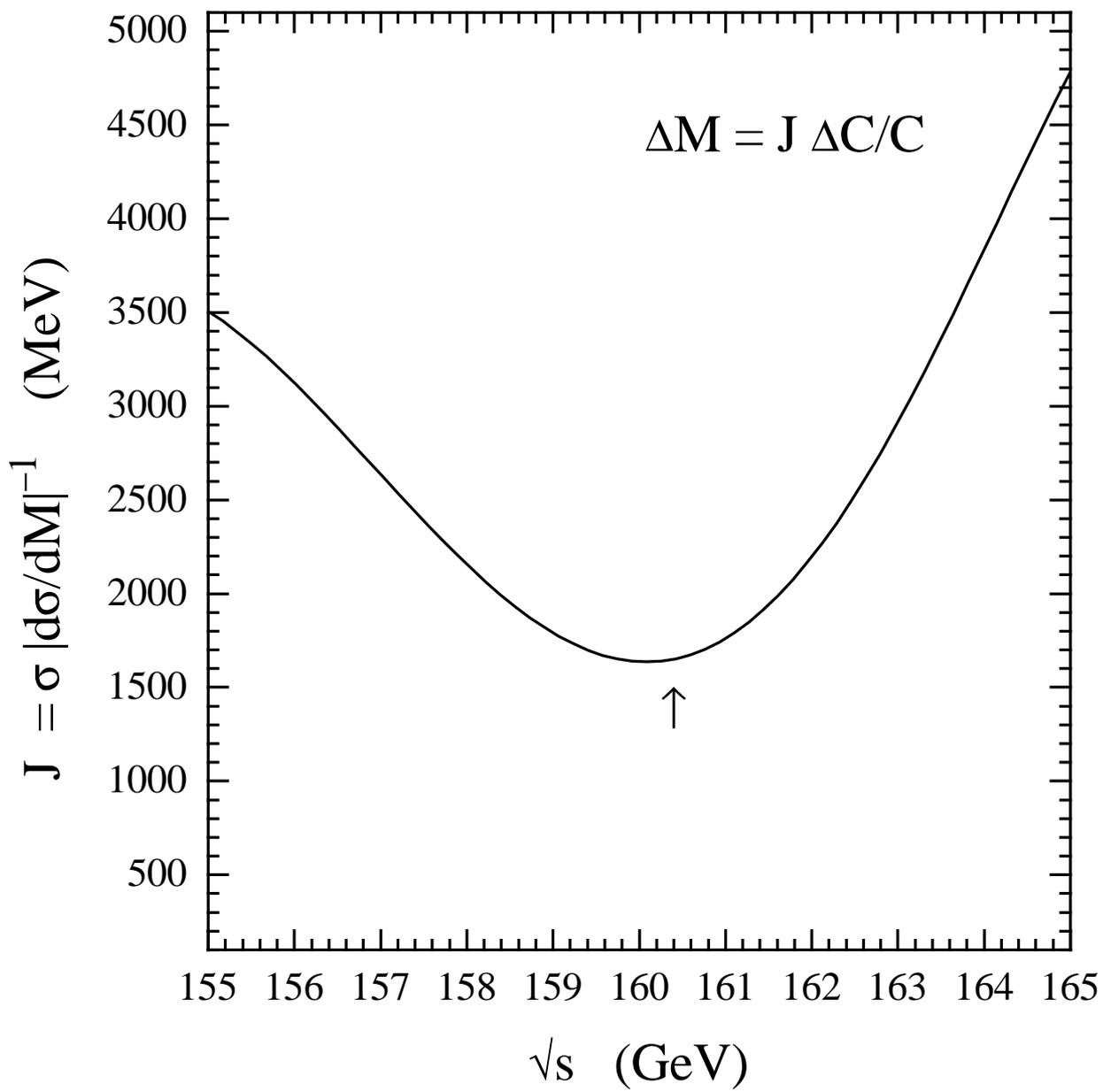
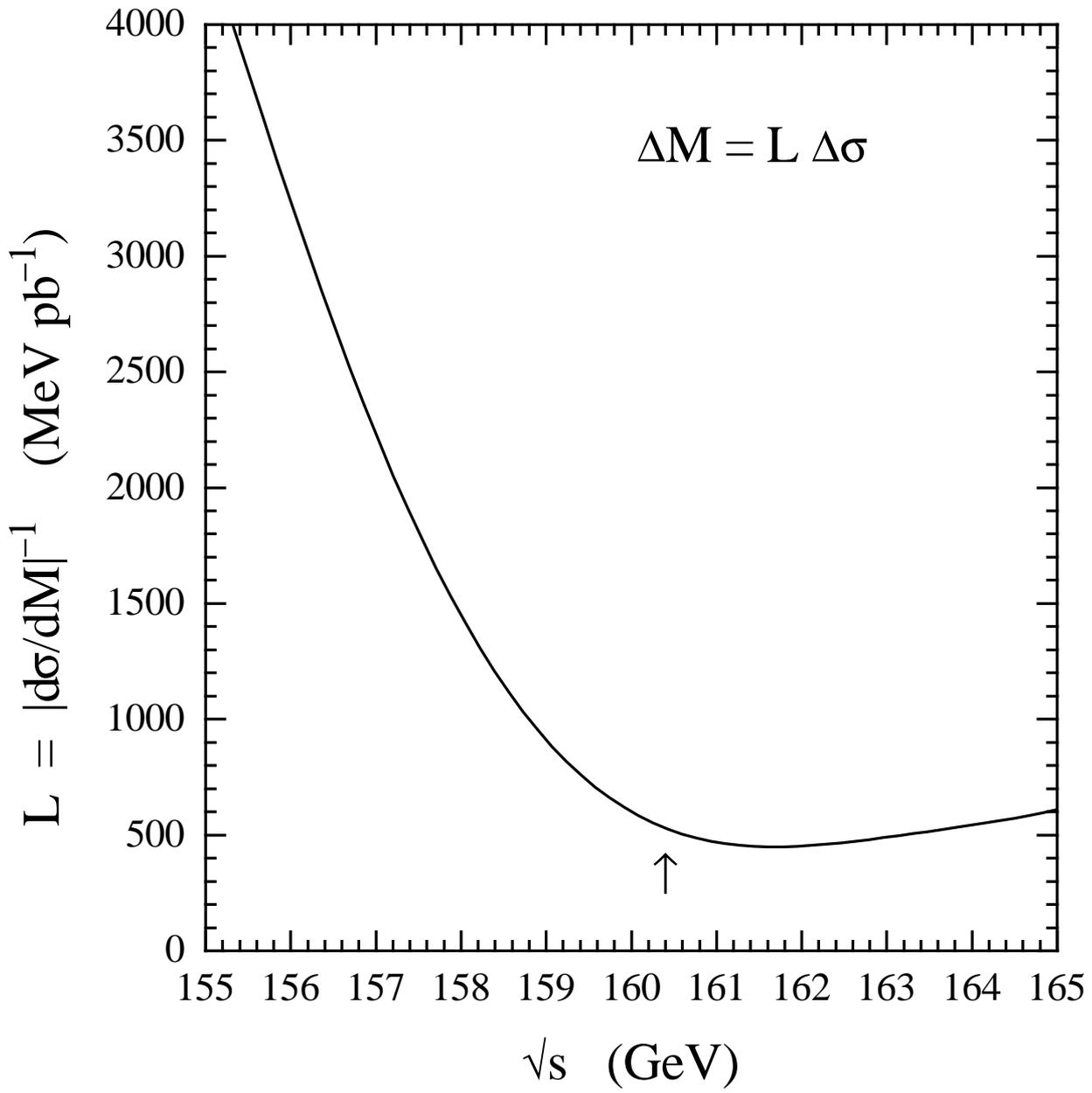
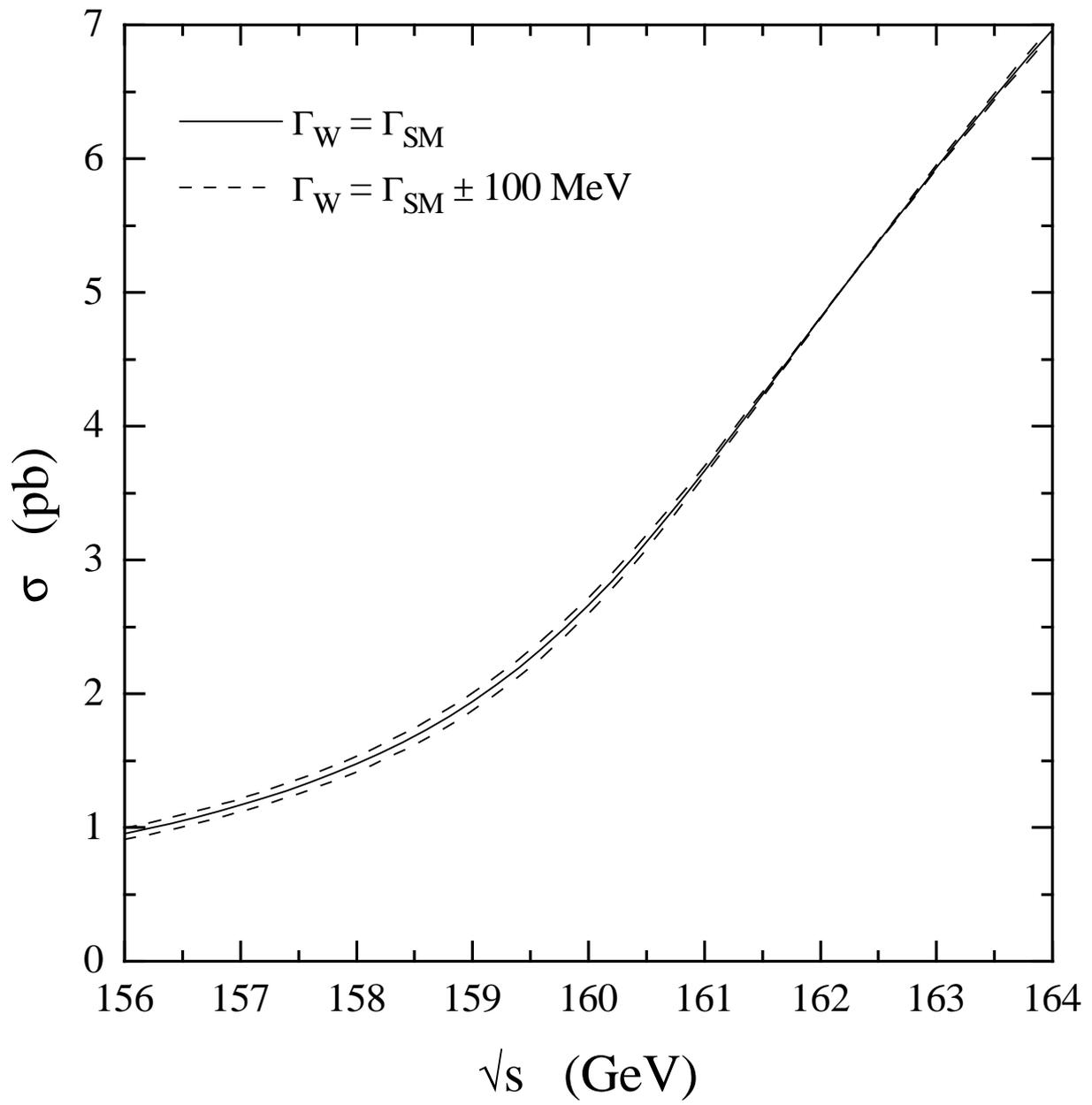
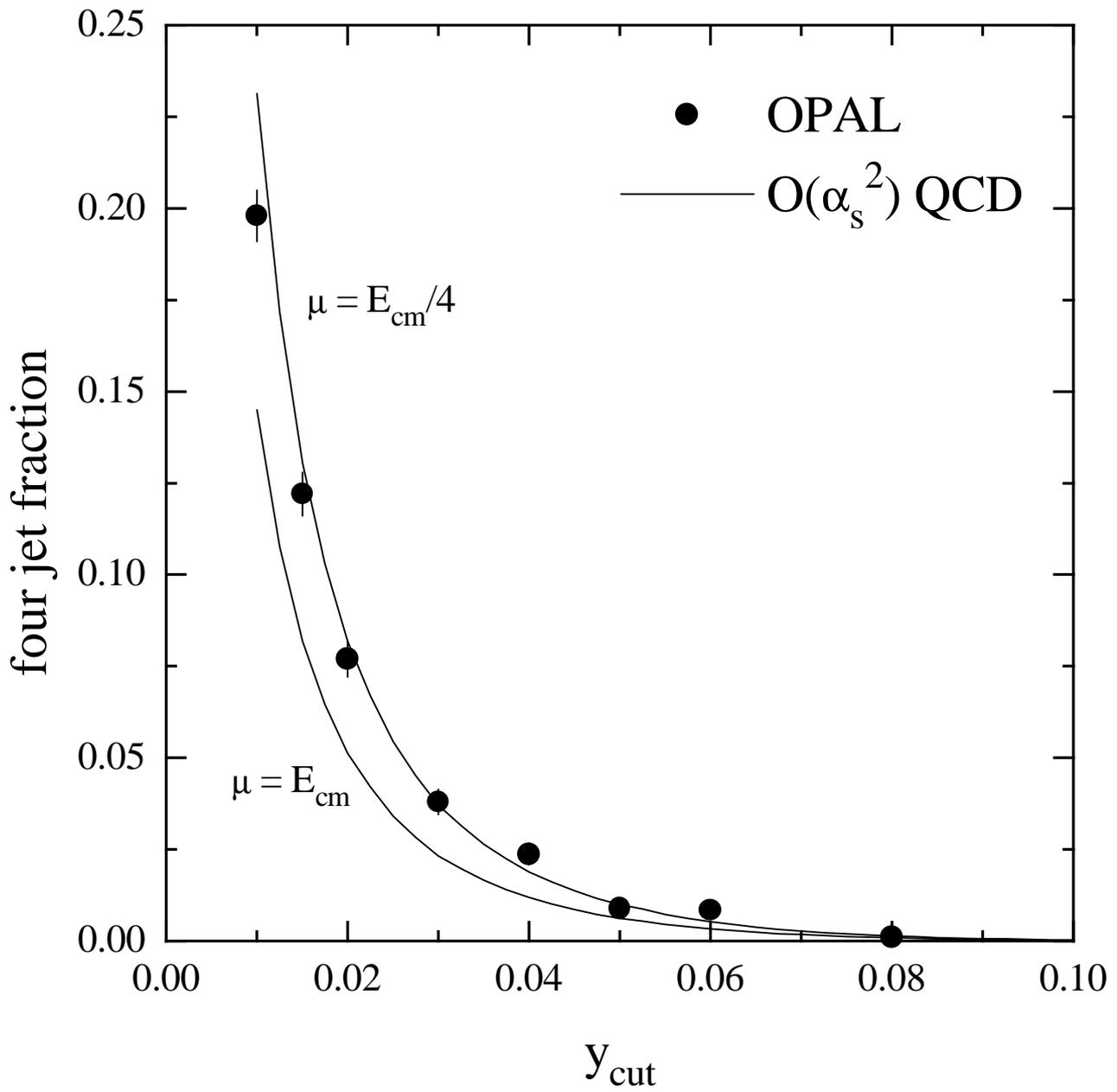
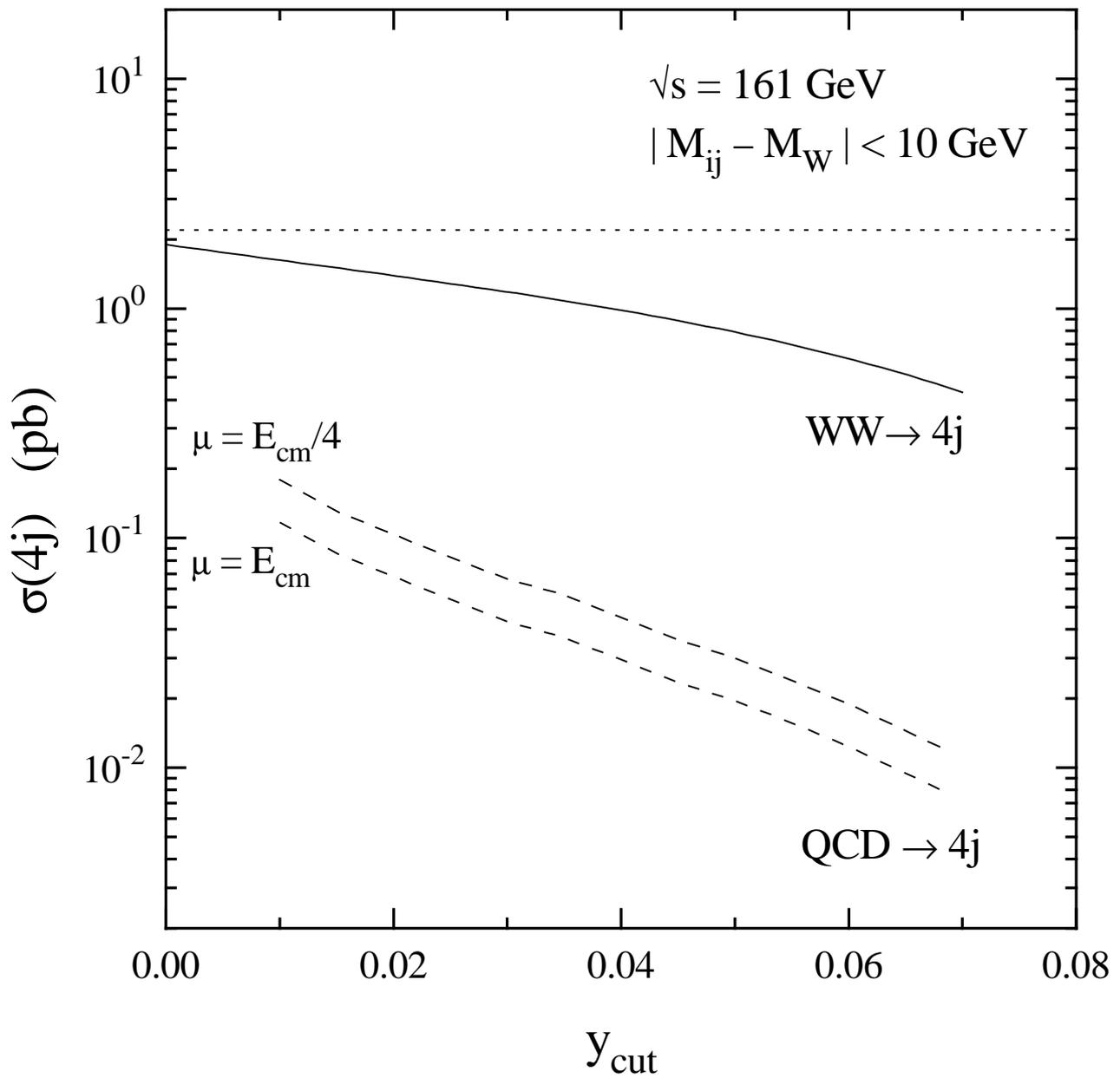
\end{document}